\documentclass[sigconf]{acmart}
\AtBeginDocument{%
  }

\setcopyright{acmlicensed}
\copyrightyear{2026}
\acmYear{2026}
\acmDOI{XXXXXXX.XXXXXXX}
\acmConference[Models '26]{Make sure to enter the correct
  conference title from your rights confirmation email}{October,
  2026}{Malaga, Spain}

\acmISBN{978-1-4503-XXXX-X/2018/06}

\usepackage{subcaption}
\usepackage{amsthm}
\usepackage{algorithm}
\usepackage{algpseudocode}
\usepackage{amsmath}
\newtheorem{exa}{Example}

\newcommand{\cC}{\mathcal{C}}

\newcommand{\cL}{\mathcal{L}}

\begin{document}

\title{Category-Based MLM: Unifying Powertypes with Superclasses}


\author{Mira Balaban}
\email{mira@bgu.ac.il}
\affiliation{%
  \institution{Ben-Gurion University of the Negev}
  \city{Beer Sheva}
  \country{Israel}
}

\author{Azzam Maraee}
\email{azzamm@live.achva.ac.il}
\affiliation{%
  \institution{Achva Academic College}
  \city{Beer Tuvia}
  \country{Israel}}

\author{Arnon Sturm}
\orcid{1234-5678-9012}
\email{sturm@bgu.ac.il}
\affiliation{%
  \institution{Ben-Gurion University of the Negev}
  \city{Beer Sheva}
  \country{Israel}
}

\renewcommand{\shortauthors}{Balaban et al}
\begin{abstract}
\textit{MultiLevel software Modeling} (\textit{MLM}) suggests that conceptual modeling in broad subject domains might require abstraction of multiple classification levels. 
The MLM approach relies on philosophical arguments, claiming  
that faithful modeling of real-world domains involves repeated type classification as in ontologies of natural kinds. 
MLM leveled architecture 
is interwoven and defined by \textit{instance-of} interlevel relationships between \textbf{\textit{clabject}}  classes in lower levels to classes termed \textbf{\textit{category}} classes, in upper levels. The \textit{instance-of} relation denotes membership of clabjects as type objects in their (powertypes) category classes, and is not transitive. 

All MLM approaches support forms of \textit{deep characterization}, i.e., category classes can influence classes in lower levels.
Deep characterization is an essential feature of  superclasses, and contradicts the non-transitive membership meaning of \textit{instance-of}.

In this paper, we introduce the \textit{Category-Based MLM} (\textit{CatMLM}) model, in which category classes have \textbf{\textit{dual superclass and powertype facets}}, based on distinction between \textit{category features} that do not participate in deep characterization, to \textit{object features}, that do. 
This distinction clarifies the role of levels, provides a clear quantifiable criterion for leveling, and yields a decision rule between the subclass and \textit{instance-of} relations.


The contribution of this paper is in introducing a well-defined MLM model that 
(1) is based on simple, quantifiable level decisions; 
(2) clarifies how leveling emerges from domain needs; and
(3) analyzes gains and losses of MLM vs. plain OO modeling.

\end{abstract}



\maketitle

\section{Introduction} 

\label{section:introduction}

\textit{MultiLevel software Modeling} (\textit{MLM}) suggests that conceptual modeling in broad subject domains might require abstraction of multiple classification levels. 
The MLM approach relies on philosophical arguments, claiming  
that faithful modeling of real world domains involves repeated type classification as in ontologies of natural kinds~\cite{henderson2012mathematics}.

This intention is captured by the \textit{PowerType analysis pattern}~\cite{Odell94,gonzalez2006powertype},
which suggests: (1) to associate a superclass with a  \textit{PowerType class}; and (2) to relate the subclasses of the superclass to the \textit{powertype class}, using \textit{instance-of} relationships. The intention is to characterize the subclasses as \textit{type objects} of the powertype class.

MLM, as a distinguished modeling paradigm, started more than twenty years ago~\cite{atkinson1997meta,atkinson2001essence}, along with the rise of the UML and the metamodeling approach in software engineering. 
Later, MLM ideas were supported by engineering arguments about accidental complexity~\cite{atkinson2008reducing,AC-MLM2021}.
MLM has triggered the development of quite a few theories, e.g.,~\cite{Carvalho2018,rossini2012graph,Neumayr2016,Bal-Khi-Kif-Mar-2018,Hinkel19}, and there are several MLM environments and applications \cite{deLara:2010:metadepth,frank2014multilevel,JeusfeldN16,flexiblemelanee16,Guizzardi-types-taxonmies-2021,Igamberdiev2018,balaban2022mediation,somogyi2022playground}.
MLM also  initiated research on design and applicability patterns~\cite{Lara:2014:UMM:2702120.2685615}, and analysis of relationships to Domain Specific Languages (DSLs)~\cite{de2013model}.

\paragraph{\textbf{\textit{MLM concepts}}}
MLM relies on two underlying inter-related notions: \textit{\textbf{instance-of}} relationships between classes and \textit{\textbf{level}} classification of the overall conceptual model. 
\textit{Instance-of} relationships between classes imply leveling, i.e., the related classes reside on different levels, and leveled architecture interlevel \textit{instance-of} relationships.
Yet, there is no general agreement on the status of levels in MLM~\cite{Balaban2018MultilevelMW,AtkinsonKuhne2018StoryOfLevels}. 

The \textit{instance-of} relation extends standard OO modeling with a relation whose intended meaning is set membership between two classes: A \textit{member class} which  functions as a \textit{type object} in the other class, which functions as a \textit{conceptual category}.
The member class was termed \textit{\textbf{clabject}}, i.e., a \textbf{\textit{cla}}ss which is also an o\textbf{\textit{bject}}~\cite{atkinson2000meta,atkinson2001essence}. 
We suggest using the term \textbf{\textit{category}} for the parent (containing) class. That is, \textit{instance-of} relationships identify $\langle clabject , category \rangle$ pairs. 


%

\paragraph{\textbf{\textit{Superclasses, powertypes, and Deep characterization}}} 
In standard software modeling, subclassing (specialization) relation is transitive, implying that subclasses inherit all features of their ancestors, implying that instances of subclasses satisfy all ancestor invariants. This property implies the Liskov Substitution Principle~\cite{liskovWing1994} and type safety of client visibility. 

In contrast, the \textit{instance-of} relation, which governs level interrelationships in MLM, is not transitive, and subclassing is not transitive over \textit{instance-of}\footnote{
  \textit{Instance-of} is not transitive: 
  $~~~  S \in T,~ T \in U \nRightarrow S \in U  $;
  subclassing is not transitive over \textit{instance-of}:
  $~~~ S \subseteq T,~ T \in U  \nRightarrow S \subseteq U $ 
    }.
Therefore, it is incorrect to infer that 
instances of clabjects are instances of their category classes, or that subclasses of clabjects are subclasses of their category classes. Hence, clabjects, in their class facet (i.e., as classes), do not inherit the features of their category classes.

Nevertheless, all MLM approaches support different forms of \textit{deep characterization}, i.e., category classes influence their \textit{offspring}\footnote{
    The \textit{offspring} relation is the transitive closure of the union of the \textit{instance-of} and subclass relations.
    } 
classes, i.e., classes in lower levels that are related by chains of \textit{instance-of} and subclass relationships~\cite{atkinson2001essence,deLara:2010:metadepth,de2013model,neumayr2014dual,frank2022multi,Igamberdiev2018,MLM-USE-MULTI2024,JeusfeldN16}.  

Deep characterization is an essential feature of superclasses: Influencing their subclasses via full inheritance of features. 
Hence, we realize that in practice, \textit{\textbf{a category class behaves like a superclass with respect to the features that participate in its deep characterization impact}}. This observation implies the underlying assumption of Category-based MLM: \\
MLM category classes have \textit{dual superclass and powertype facets} (following the superclass and its associated powertype class, in the \textit{Powertype} analysis pattern). Accordingly, the features of a category class are classified into:
\begin{enumerate}
  \item 
    \textit{\textbf{Object features}}, i.e., the features that participate in the deep characterization mechanism, and apply to data objects of offspring classes along the levels hierarchy. They define the \textit{superclass facet} of the class.  
  \item
    \textit{\textbf{Category features}}, i.e., features that characterize instances of the category class alone (no inheritance). They define the \textit{powertype facet} of the class.
  \end{enumerate}

\paragraph{\textbf{\textit{Level characterization}:}}
The distinction between category to object features clarifies the \textit{\textbf{role of levels}} in MLM: \textit{\textbf{Introduce abstraction which is characterized by the category features of category classes}}. 
It also provides a decision rule between the subclass and \textit{instance-of} relations: Adding a category class (using \textit{instance-of}) is justified by category features, i.e., a new level of abstraction. 
The overall line of thought is: 
\begin{itemize}
  \item
    category features mark introduction of higher abstraction \\
    $\Rightarrow$
  \item
    they participate in \textit{instance-of} relationships, with their \\
    $\langle clabject, category \rangle$ pairs of classes \\
    $\Rightarrow$
  \item 
    introduction of new levels.
\end{itemize}

\smallskip
In this paper, we introduce the \textit{Category-Based MLM} (\textit{CatMLM}) model, which is based on the \textbf{\textit{dual superclass and powertype facets}} of category classes, and provides a clear quantifiable criterion for leveling. This distinction enables extension of the abstraction (encapsulation, client visibility) property of class hierarchies, which is captured by the superclass facet of category classes. 

CatMLM clarifies the status of the \textit{instance-of} relation in terms of feature influence between classes: While \textit{subclass} provides full inheritance of features and associations provide none, \textit{instance-of} provides partial inheritance, i.e., that of the object features. 

\begin{figure*}[t]
  \centering
  \includegraphics[width=\linewidth]        
                  {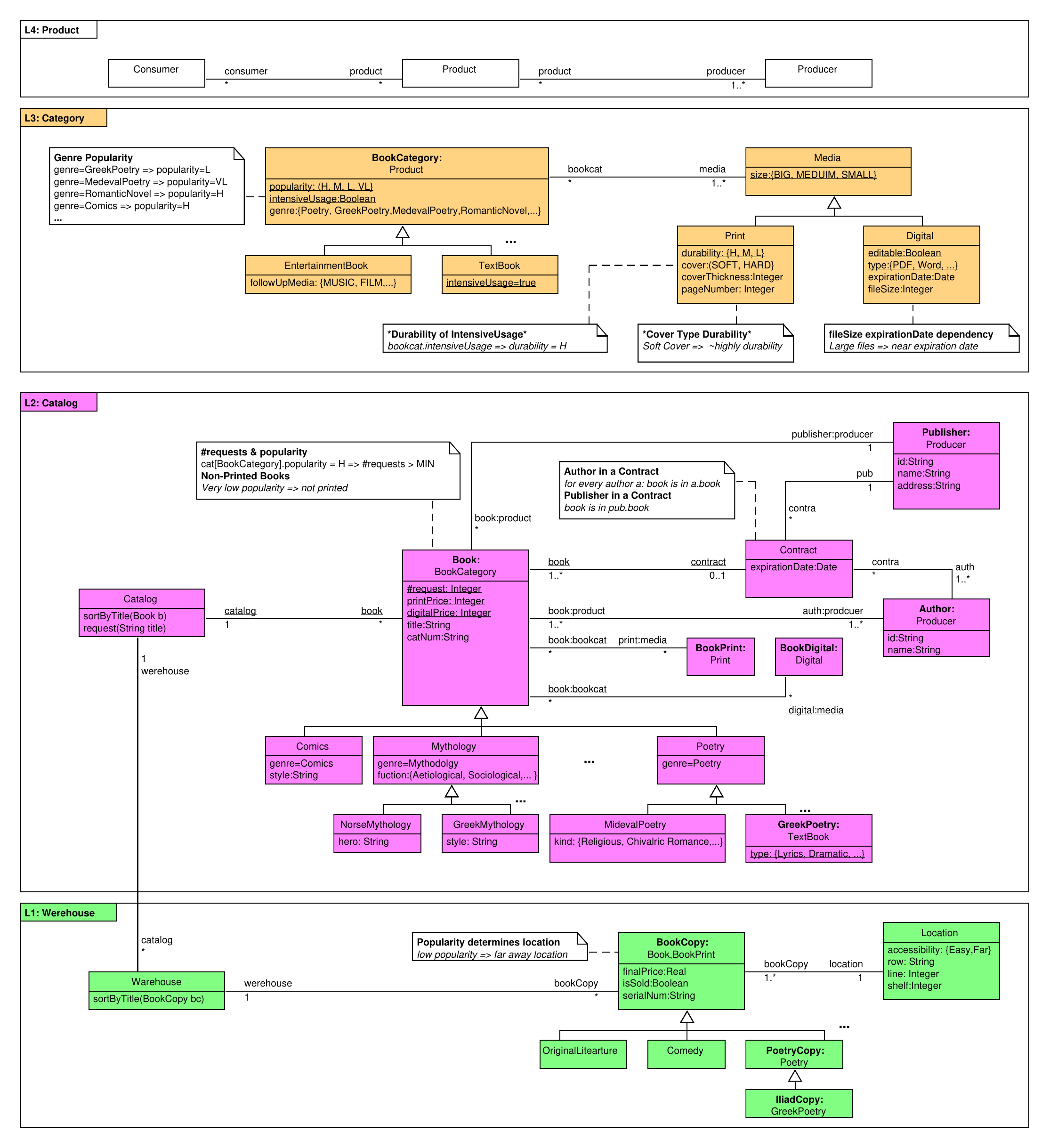} 

    \caption{A partial CatMLM model for a book warehouse enterprise} 
  \label{figure:warehouse-small}
\end{figure*}

We have developed extensive CatMLM models for two challenge MLM problems: (1) the \textit{Warehouse} and (2) the \textit{Process} problems. 
Part of the Warehouse CatMLM model is presented and intuitively explained in the following example. 
\begin{exa} [Category-based-MLM model for supporting a \textit{Warehouse} enterprise\footnote{
  One of the published MLM challenge problems concentrates on MLM for warehouse modeling~\cite{whchallenge}.}
  ]
\label{example:warehouse}
~ \\  
Figure~\ref{figure:warehouse-small} shows part of a CatMLM model that supports a warehouse enterprise that focuses on book management. This Warehouse handles hard-copy and digital books. 
The warehouse maintains a possibly in-house book-catalog, and consults a general source on book categories.
For that purpose, the warehouse keeps a knowledge model on different abstraction levels.

This MLM model consists of four levels: 
\texttt{Warehouse, Catalog, Category} and \texttt{Product}.
\begin{description} 
  \item [Warehouse]
     This bottom level describes the physical warehouse that stores hard copies of books.
  \item [Catalog]
    describes the books in the catalog, and covers catalogic features like \emph{print} and \emph{digital prices}, \emph{genre}, and \emph{number of requests}, which do not characterize book copies. The \texttt{catalog} serves as the entry point of the enterprise.
  \item [Category]
    includes general information on book categories, like book genres, their popularity, reading insensitivity, and their media characterization.
  \item [Product]
    The most general level is not yet developed.
\end{description}

Each level is visualized by a class diagram. \emph{Instance-of} (interlevel) relationships between clabjects to categories are marked  as in "$Book : BookCategory$" on level \texttt{Catalog}, with \emph{Book} being the clabject and \emph{BookCategory} being the category on the \texttt{Category} level. Note that \emph{instance-of} relationships do not necessarily coincide with exact level ordering, as in $~~Publisher : Producer~~$ on the \texttt{Catalog} level.

Category features of category classes are marked using an underline. In class \emph{Book} on the \texttt{Category} level, \emph{\#request} is a category attribute, \emph{contract} is a category role, and \emph{requests\&popularity} is a category constraint.  
In addition, the model includes an interlevel association between the \emph{Warehouse} and the \emph{Catalog} classes.
%
\end{exa}

CatMLM is a formal MLM model that has the following characteristics: 
(1) is based on simple, quantifiable level decisions; 
(2) clarifies how leveling emerges from domain needs; 
(3) extends the abstraction modeling of class hierarchies with the superclass facet of category classes.
Furthermore, the simple quantifiable nature enables the development of pragmatic guidelines.

CatMLM simplifies and improves our previous \textit{Mediation-based MLM theory} (\textit{MedMLM})~\cite{Bal-Khi-Kif-Mar-2018}\footnote{
  MedMLM implementations include FOModeler~\cite{balaban2022mediation} and the MLM-USE tool~\cite{MLM-USE-MULTI2024,MLM-USE-2025} (which is built on top of the USE modeling tool~\cite{use-tool-2007,use_github}).
  }.
The complex interlevel \textit{mediators} of MedMLM are replaced by the simple and conceptually justified distinction between \textit{category} to \textit{object} features.   
CatMLM is currently being implemented on top of the MLM-USE tool.


Section~\ref{section:category-based-MLM} defines the CatMLM model, and
section~\ref{secrion:catMLM related} discusses and compares MLM features of CatMLM and other related MLM approaches.
Section \ref{section:conclusion} concludes the paper and summarizes plans for the future.



\section{Category-based MLM}
\label{section:category-based-MLM}

The underlying motivation behind MLM is the understanding that in addition to standard OO class-hierarchy abstraction, there is a broader abstraction by families of classes of elements. 
This is the trigger behind the powertype design pattern and similar ontological approaches~\cite{Odell94,gonzalez2006powertype,henderson2012mathematics}.
In other words, the role of the category classes, which are introduced by \textit{instance-of} relationships, is to introduce higher levels of abstraction, that does not exist, 
and is not relevant in the lower levels of their clabjects. Therefore, a category class 
\textit{\textbf{must have features that are not relevant to its clabjects, and are not inherited}}. 

Nevertheless, we know that all MLM approaches employ some form of deep characterization, i.e., their features influence the features of their offspring classes in lower levels. 
Moreover, in most cases, the influence is achieved by plain inheritance. In other words, a 
category class carries features that characterize the objects of their clabjects, i.e., \textit{\textbf{category classes function as superclasses of their clabject classes}}. 

The underlying concepts of CatMLM are: \textit{level, instance-of} that relates a \textit{Category} and a \textit{Clabject} classes, the distinction between \textit{category to object features} of category classes, and its impact on deep characterization semantics.

\paragraph{\textbf{\textit{CatMLM --- syntax}}}

A model in Category-based MLM is a collection of \textit{MLM-dimensions}, where an \textit{MLM-dimension} is a total order of \textit{levels}, i.e., a \textit{level-dimension}: $\cL_1, \ldots, \cL_n$, where $\cL_1, \cL_n$ are the \textit{bottom} and \textit{top} levels, respectively.
A more general view might define a CatMLM model as a partially ordered set of \textit{levels}. In practice, all MLM approaches are restricted to a single dimension. Henceforth, we continue with the single level-dimension view.\footnote{
    There is a limited discussion about level-dimensions, where a model consists of multiple totally ordered sets of levels~\cite{kuhne2025supporting,kuhne2022multi}
    } 

\smallskip
\noindent
\textbf{\textit{Levels} and \textit{instance-of}:} A \textit{level} is a plain \textit{class model}. The essence of MLM, in all approaches, is the extension of OO modeling with the \textit{instance-of} relation, which is a binary relation between classes in different levels, $\cL_i, \cL_j$, $i<j$. \\
\textbf{Notation:} In the text we use the notation $~ C^{(i)} \prec D^{(j)}$, where $C^{(i)} \in \cL_i,~ D^{(J)} \in \cL_j$, and in the visualization as in Figure~\ref{figure:warehouse-small}, it is written: $~ C^{(i)} : D^{(j)}$.~
Class $C^{(i)}$, in the lower level is the \textit{clabject}, and class $D^{(j)}$ in the upper level  is the \textit{category} class. \\
In the CatMLM model in Figure~\ref{figure:warehouse-small}, there are four levels: \\
\textbf{\textit{1})} On level \texttt{Warehouse} there are three clabjects: \\
$BookCopy \prec Book, BookPrint$, 
$~PoetryCopy \prec Poetry$, 
$~IliadCopy \prec GreekPoetry$.
All category classes are on level \texttt{Catalog}. \\
\textbf{\textit{2})} 
On level \texttt{Catalog} there are four clabjects: \\ 
$Book \prec BookCategory$, $~GreekPoetry \prec TextBook$, 
$~Publisher \prec Producer$, 
$~Author \prec Producer$.
$~BookCategory$ and $TextBook$ are category classes on level \texttt{Category}, and $Producer$ 
is a category class on level \texttt{Product}. \\
\textbf{\textit{3})} On level \texttt{Category} there is  
a single clabject $BookCategory \prec Product$, where $Product$  is a category class on level \texttt{Product}.

Between every pair of adjacent levels, $\cL_i, \cL_{i+1}$, there is at least one pair of \textit{instance-of} related $\langle clabject, category\rangle$ classes.
That is, \textit{instance-of} is restricted to relate classes in different levels, and level distinction is marked by existence of \textit{instance-of} relationships.

The question of exact synchronization between \textit{instance-of} and level ordering has been raised in multiple approaches. Our experience with the two challenge examples that we have developed~\cite{balaban2025pragmatic} shows that this is too restrictive. In the Book Warehouse example in Figure~\ref{figure:warehouse-small}, indeed the \textit{instance-of} classification of \textit{Publisher} and \textit{Author} on level \texttt{Catalog} as clabjects of category \textit{Producer} on level \texttt{Product} crosses over level \texttt{Category}.

\smallskip
\noindent
\textbf{\textit{Object} and \textit{category} features:}
Category classes carry two kinds of features (attributes, roles, constraints): \textit{Object features} and \textit{Category features}. Object features are plain OO features, and therefore, obey the inheritance semantics of OO modeling.  
Category features are relevant only for objects of the category classes. 

Roles are components of associations. Therefore, the object/category dichotomy must be defined for associations. In CatMLM, associations that involve category classes can be classified as object or category ones. In an object or category association, both roles are object or category roles, respectively. This syntactic constraint is in effect for all associations, including associations where only one class is a category.   

\begin{figure*}[t]
  \centering
  \includegraphics[width=\linewidth]        
                  {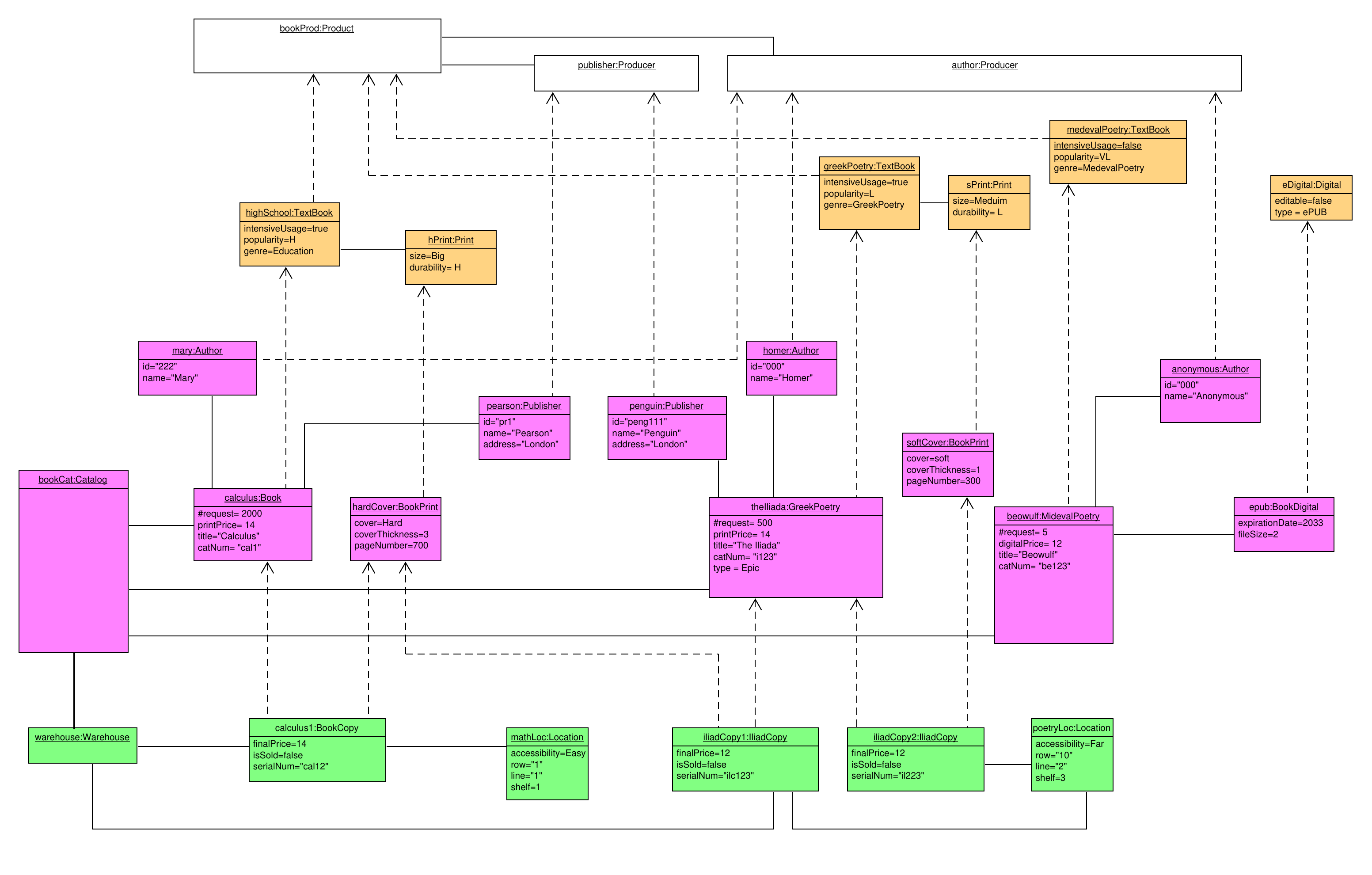} 
    \caption{A legal instance of the CatMLM model in Figure~\ref{figure:warehouse-small}}
    \label{figure:warehouse-object}
\end{figure*}

In the visualization in Figure~\ref{figure:warehouse-small}, category features are underlined while object features are not: 
\begin{enumerate}
   \item
     \textbf{Attributes:} 
     Category class \textit{Book} on level \texttt{Catalog}, has three \textit{category attributes} (not inherited) and two \textit{object attributes} (inherited by \textit{BookCopy }on the \texttt{Warehouse} level)
  \item 
    \textbf{Roles:}
    The $\langle bookcat, media \rangle$ association on level \texttt{Category} is an \textit{object association} (object typed). Its roles are inherited by clabjects of \textit{BookCategory} and \textit{Media} on level \texttt{Catalog}.  The inherited association is visualized, for further clarity, but could be avoided (unless the role renaming is important). 
    
     The $\langle book, format \rangle$ association on level \texttt{Catalog} is a \textit{catalog association} (catalog typed). Role \textit{format} is not inherited by clabjects of \textit{Book} on level \texttt{Warehouse}.

    
  \item 
    \textbf{Constraints:}
    \textit{Object constraints} are ones that access only object attributes or roles. For example, the "\textit{fileSize \textit{expirationDate} dependency}" that is imposed on class \textit{Digital}, i.e., "\textit{Digital} typed" (its signature is $Digital \rightarrow Boolean$), in level \texttt{Category}. Such constraints are inherited by clabjects, via the \textit{instance-of} relation as in standard OO class hierarchy.

    \textit{Category constraints} (underlined) are ones that apply only to instances of their category class. For example, the "\textit{$\#$requests $\&$ popularity}" constraint that is \textit{Book} typed, in level \texttt{Catalog}, applies only to \textit{Book} objects, but not to objects of its clabject \textit{BookCopy}. 
    
    \textit{Object-category constraints} are category constraints that a modeler might mark as relevant to clabject objects.
    For example, the "\textit{Cover Type Durability}" constraint that is \textit{Print} typed on level \texttt{Category}, refers to category attribute \textit{durability} (underlined). 
    Inheritance of such constraints relies on application-specific machinery for linking clabject objects with their category objects. In CatMLM, these are links of the relevant \textit{instance-of} relation. For example, in Figure~\ref{figure:warehouse-object} objects \textit{iliad1, iliad2} are linked to object \textit{theIliad}, of category class \textit{GreekPoetry} on level \texttt{Catalog}.
\end{enumerate}

\smallskip
\noindent
\textbf{Interlevel elements:} 
A CatMLM model might include interlevel associations and interlevel constraints. Both apply to objects of different levels. Interlevel associations characterize relationships between objects of different levels, and interlevel constraints impose restrictions on such objects.

Altogether there are three kinds of associations/constraints: (1) Category associations/constraints at various levels, that apply, each, on instance objects of the levels they are defined at; (2) Object associations/constraints at various levels, that are inherited along offspring chains, and apply to instances at multiple levels; (3) Interlevel associations/constraints, that restrict overall object instances of a CatMLM model. 

\smallskip
\noindent
Syntax constraints on CatMLM models: \\
\textit{\textbf{1)} Leveling requires \emph{instance-of}}: 
A level must have a class that participates in an \textit{instance-of} relationship (as a clabject or as a category class). \\
\textit{\textbf{2)} Category classes}: A category class must have a category feature

\paragraph{\textit{\textbf{CatMLM --- Semantics}}}

The semantics of MLM models is not explicitly discussed in the MLM community. In many MLM examples, object population of the bottom level is attached to the model and  marked as Level 0.
The message seems to be that an MLM model  denotes legal object models of Level 1, and that in MLM, models and data models are unified.

Indeed, in the MLT, MLT* effort~\cite{Carvalho2018,AlmeidaFC17,fonseca2021multi}, which introduces an axiomatic specification of a leveled (ordered) world of types, all individual (data) objects reside on the bottom level (order), and levels are organized by \textit{instance-of} interlevel relationships (MLT* enables cross-level \textit{instance-of} relationships).

Such MLM models raise major questions about the intended semantics of classes in the model. Mainly,
\begin{quote}
  \textit{What is the object denotation of classes in middle levels?} \\
  \textit{What is the link denotation of interlevel associations?}
\end{quote}
In particular, what is the object denotation of category classes? For a category \textit{C}, its objects are: (1) only the clabjects in the model? (2) can there be objects outside the MLM model? (3) are all objects of \textit{C} type-objects (clabjects)? 

\smallskip
\noindent
\textbf{Observation 1:} In real MLM models like the Warehouse example in Figure~\ref{figure:warehouse-small},  \textit{classes in internal levels, including category classes, must have their own objects}. 
For example, the \textit{Book} category class on the \texttt{Catalog} level captures an abstraction of all books in the catalog. These books are objects of the \textit{Book} class, and they exist independently of the objects of the clabject \textit{BookCopy} on the \texttt{Warehouse} level. 
Likewise, the \textit{Author} non-category class on that level, has its own objects.
In particular, \textit{category classes must enable multiple values for their category features}. 

\smallskip
\noindent
\textbf{Observation 2:}
In real MLM models, like the Warehouse example in Figure~\ref{figure:warehouse-small}, data objects of clabject classes must be able to retrieve the specific category information that is relevant for them. 
For example, in Figure~\ref{figure:warehouse-object}, objects of \textit{BookCopy} (green) must retrieve their specific \textit{Book} features (purple). \textit{Calculus1} (green) retrieves its \textit{Book} category values from \textit{calculus} (purple), and its \textit{BookPrint} category values from \textit{hardCover} (purple); \textit{iliad1, iliad2} (green) retrieve their \textit{Book} category values from \textit{theIliada} (purple), and diverge on their \textit{BookPrint} category values, as specified by \textit{hardCover} and \textit{softCover} (purple) .  

\smallskip
\noindent
\textbf{Conclusions:}
\begin{enumerate}
  \item
    Classes on internal levels, including category classes, must have plain objects as their data population.
  \item 
    Objects of clabjects must be able to retrieve their own (specific) category information.
\end{enumerate}


\medskip
CatMLM follows these conclusions. 
An \textit{instance of a CatMLM model} might include instance objects of classes at every level, including category classes (first conclusion). Moreover, objects of clabject classes must be linked by \textit{instance-of} relations to objects of their category classes, one object for each category class.  
Therefore, a \textit{legal instance} of a CatMLM model consists of \textit{legal instances of the class models} of the levels, that (1) satisfy all inter-association multiplicities; (2) satisfy all constraints; (3)  every object of a clabject is linked via \textit{instance-of} to an object of each of its category classes. 

In many cases the \textit{intended instances} of a CatMLM model consist of legal instances of the bottom level, which are extended with objects of other levels, following inter-associations and \textit{instance-of} relations. Example~\ref{example:warehouse-instance} presents such an instance. 
\begin{exa}
    [A legal intended instance of the CatMLM model in Figure~\ref{figure:warehouse-small}]
\label{example:warehouse-instance}
\end{exa}
\vspace{-1mm}
\noindent
Figure~\ref{figure:warehouse-object} shows an intended object instance of the CatMLM model in Figure~\ref{figure:warehouse-small}.
This instance concentrates on six objects of classes in the bottom \texttt{Warehouse} level (green color), and extends them with interlinked objects through mandatory inter-associations (solid lines) and \textit{instance-of} relations (dashed lines). 
\begin{enumerate}
  \item 
    \textbf{Objects:} \\
    The instance of the \texttt{Warehouse} level includes a \textit{Warehouse} object, three \textit{BookCopy} objects and two \textit{Location}s. \\
    The \texttt{Catalog} level instance includes three\textit{Book} objects, each with its \textit{Author} and \textit{Publisher}, two \textit{BookPrint} objects, and one \textit{BookDigital} object. Note that the \textit{beowulf} object of \textit{MideavlPoetry} has a \textit{BookDigital} format, and does not serve as a category object for any \textit{BookCopy} (because the Warehouse carries only printed copies).  \\
    The instance of the \texttt{Category} level includes three \textit{BookCategory} objects, and three \textit{Media} objects. \\
    The instance of the \texttt{Product} level includes one \textit{Product} and two \textit{Producer}s.
  \item
    \textbf{Inter-association link:} Object \textit{warehouse} of class \textit{Warehouse} must be linked to the single object of the \textit{Catalog} class (purple color), due to the inter-association between \textit{Warehouse} and \textit{Catalog}. 
  \item 
    \textbf{\textit{Instance-of} interlinks of \texttt{Warehouse} level objects:} \\
    Object \textit{calculus1} (green) of class \textit{BookCopy} is linked via \textit{instance-of} to object \textit{calculus} (purple) of \textit{Book} and to object \textit{hardCover} (purple) of \textit{BookPrint}. \\
    Objects \textit{iliadCopy1, iliadCopy2} (green) of class \textit{IliadCopy} are linked via  \textit{instance-of} to object \textit{theIliada} (purple) of class \textit{GreekPoetry}, and  diverge on their links to  \textit{hardCover} and \textit{softCover} (purple).
  \item 
    \textbf{\textit{Instance-of} interlinks of \texttt{Catalog} 
    level objects:} \\
    Object \textit{calculus} (purple) is linked via \textit{instance-of} to object \textit{highSchool} (beige) of category class \textit{TextBook}. 

    Object \textit{theIliada} (purple) is linked to object \textit{greekPoetry} (beige) of category class \textit{TextBook}. 

    Objects \textit{hardCover} and \textit{softCover} (purple) are linked to objects \textit{hPrint} and \textit{sPrint} (beige) of category class \textit{Print}, respectively.

    The four objects (purple) of the \textit{Publisher} and \textit{Author} clabjects are linked to twoobjects of category class \textit{Producer}, two levels above. 
  \item
    \textbf{Constraints:} 
    \begin{enumerate}
      \item
        "\textit{Popularity determines location}", which is imposed on objects of \textit{BookCopy} in level \texttt{Warehouse}: The constraint states that copies of less popular books should be located in a less accessible place. 
        \textit{calculus1} retrieves its \textit{popularity} value (a category feature) via the \textit{instance-of} path of links: \textit{calculus1 --> calculus -->highschool}. The popularity value is "\texttt{H}", and indeed, \textit{calculus1} is located in \textit{mathLoc1}, whose \textit{accessibility} is \texttt{Easy}.

        \textit{iliadCopy1, iliadCopy2} retrieve their \textit{popularity} value (a category feature) via the \textit{instance-of} path of links: \textit{iliadCopy1/2 --> theIliada -->greekPoetry}. The popularity value is "\texttt{L}", and indeed, \textit{iliadCopy1, iliadCopy2} are located in \textit{poetryLoc1}, whose \textit{accessibility} is \texttt{Far}.
      \item 
        "\textit{\#requests\&popularity}", which is imposed on objects of \textit{Book} in level \texttt{Catalog}: The constraint states that the number of requests for books whose category has high popularity must exceed some known minimum.   The object \textit{calculus} retrieves its popularity value via the \textit{instance-of} link \textit{calculus-->highSchool}. The popularity value is \texttt{H} (High), and the number of requests for \textit{calculus} is $2000$, which exceeds the warehouse's minimum threshold.
   
      \item 
        "\textit{Non-Printed books}", which is imposed on objects of \textit{Book} in level \texttt{Catalog}: The constraint states that books whose category has very low popularity are not printed. The object \textit{beowulf} retrieves its popularity value via the \textit{instance-of} link 
        \textit{beowulf-->medevalPoetry}.
         The popularity value is  \texttt{VL} (Very Low), and \textit{beowulf} is available only in digital \textit{epub} format and has no physical copy in the warehouse. 
    \end{enumerate}
\end{enumerate}    
$\square$
%

\paragraph{\textbf{Computation abstraction in CatMLM}}

The different semantics of \textit{category} and \textit{object} features implies their distinct inheritance behavior. While \textit{category-typed features} of a category class are not inherited,
\textit{object-typed features} of a category class \textit{C} are fully inherited by all clabjects of \textit{C}, via their \textit{instance-of} relationships with \textit{C}.
That is, with respect to object-typed features, \textit{instance-of} behaves like subclass, and the class satisfies the \textit{Liskov substitution principle} (\textit{LSP})~\cite{liskovWing1994}. Category-typed features violate the LSP.  

This distinction affects client operations/constraints of a category class {$\cC$}. Operations or constraints that take {$\cC$} objects as arguments, i.e., {$\cC$} typed, can be classified, based on whether they access category features of \textit{$\cC$}, or alternatively, access only object-typed features of {$\cC$}. In the latter case, they can be typed as $\cC_{object}$ operations (constraints).  
Operations/constraints that are $\cC_{object}$ typed satisfy the LSP: They can take objects of offsprings of $\cC$ as arguments.  
Therefore, a category class can explicitly declare its object type $\cC_{object}$, and its clients can refine their signature, i.e., distinguish between the general type of $\cC$ to the $\cC_{object}$ type. 
This is the basis for the CatMLM type theory.
For example, in Figure~\ref{figure:warehouse-small}: 
\begin{enumerate}
  \item
    Constraint "\textit{fileSize expirationDate dependency}" on level \texttt{Category} can have the more restricted $Digital_{object}$ type;
  \item 
    The type of method $sortByTitle(Book b)$ can be refined into 
$sortByTitle($ $Book_{object} b)$.
\end{enumerate}



\section{Related Works}

\label{secrion:catMLM related}
Multi-level modeling encompasses a wide range of frameworks, tools, and applications that differ in their theoretical foundations, design objectives, and implementation strategies. Regarding model organization, most MLM approaches are limited to a single classification dimension. The multi-dimensional approaches presented in~\cite{kuhne2022multi,kuhne2025supporting} extend this structure by introducing orthogonal dimensions that support independent classification hierarchies while preserving soundness checking.

A further distinction concerns implementation strategy. Some approaches provide dedicated MLM languages and environments~\cite{flexiblemelanee16,deLara:2010:metadepth,frank2018flexible}, whereas others embed or emulate MLM concepts within established two-level modeling infrastructures~\cite{balaban2022mediation,MLM-USE-MULTI2024,MLM-USE-2025,Selway:2017:CFL:3143817.3144049}. MedMLM is implemented in FOModeLer, based on FOML, and in MLM-USE, which extends the USE environment; CatMLM is planned as an extension of MLM-USE. In contrast, Melanee, MetaDepth, and FMML$^{x}$ offer native MLM languages and environments with native support for multi-level modeling and varying degrees of support for execution and constraint specification.

\subsection{Deep Characterization in MLM}

Deep characterization refers to the capability of category classes to influence their clabject classes. This capability is not a natural feature of the \textit{instance-of} relation, and is not derived from its intended meaning. 
Yet, all MLM systems recognize this need. It has the effect of selective application of the feature inheritance mechanism, that characterizes specialization (subclass-superclass) relations. 

The dominant method of deep characterization is the method of \textit{Potency marking}, which assigns an "influence distance" to elements in a level~\cite{atkinson2001essence,deLara:2010:metadepth,neumayr2014dual,frank2022multi,Igamberdiev2018,JeusfeldN16,fonseca2021multi,atkinson2024misconceptions}. 
In principle, potency values 
can be assigned to classes, attributes, associations and constraints. 
The potency value of a class might specify the number of \textit{instance-of} relations below the class; the potency value of an attribute might specify the number of levels of inheritance of the attribute until it is given a final value; and for an association and a constraint, the potency value might specify the number of levels they are inherited.  

Multiple variants of potency-based deep characterization have been suggested and used, e.g., \textit{attribute concretization}~\cite{frank2022multi}
and \textit{leap potency}, that allows leaps in the influence of an element~\cite{de2014extending,Lara:2014:UMM:2702120.2685615}. 
Melanee supports unbounded, \emph{star}, potency, \emph{mutability}, which fixes the value-finalization level, and \emph{durability}, which lets a finalized value remain inherited ~\cite{Gerbig2016AFC,flexiblemelanee16}. SLICER provides no potency marking; deep characterization is controlled through specific relations \cite{Selway:2017:CFL:3143817.3144049}. Attributes may be inherited, refined, or remain partially uninstantiated until a value is assigned at a lower level.
Another deep characterization approach involves \textit{default/selected feature inheritance} from category classes to their clabjects~\cite{balaban2022mediation,MLM-USE-MULTI2024,MLM-USE-2025}. 

A different aspect of deep characterization involves \textit{deep constraints}, i.e., constraints that apply to elements in multiple levels. Such constraints might  employ inter-level specific keywords, as in  the constraint languages of Melanee and FMML$^x$~\cite{atkinson2015DeepConstraints,lange2023docl,frank2018flexible}.

MedMLM~\cite{Bal-Khi-Kif-Mar-2018,MLM-USE-MULTI2024,MLM-USE-2025} and CatMLM enable several forms of deep characterization.
Both models enable \textit{interlevel constraints and associations}. Interlevel associations relate classes that reside on different levels. Interlevel constraints involve elements from different levels. For example, a constraint might follow links of an interlevel association or instance-of (naturally interlevel) relations. 

MedMLM and CatMLM enable selected inheritance of features of category classes, by their clabjects. In MedMLM all features of a category class are inherited by default, unless explicitly prevented in the mediator of the level of the clabject. In CatMLM, category features are not inherited, while object features are inherited.

In addition, means for interlevel influence can be built, as derived operations, based on the category/object inheritance mechanisms of CatMLM (and similarly, in MedMLM). For example, a plain potency marking operation will take a level distance argument $k$, declare the marked feature as an object feature for $k-1$ levels, and turn it into a category feature in the $k$th level down.

\section{Conclusions and Future Research}

\label{section:conclusion}

In this paper, we have presented the CatMLM multilevel model,
which is based on the distinction of the \textit{\textbf{dual superclass and powertype facets of category classes}}. 
Based on this distinction, we introduced the distinction between \textit{category} and \textit{object features}. The \textit{category --- object} division of features is used for type refining, level structuring and as a model development principle.  
We have shown how to use  $Category_{object}$ typing for extending the computational abstraction of subclassing. 

In the future, we plan to extend our previous MLM modeling tool to fully support CatMLM observations.
In particular, we plan on the development of a formal typing theory of an \textit{instance-of} relation that supports the $Category_{object}$ and $Category_{category}$ typing distinctions.

\bibliographystyle{ACM-Reference-Format}
\bibliography{references}

\end{document}